\documentclass[12pt]{iopart}
\usepackage[T1]{fontenc}
\usepackage{graphicx}% Include figure files
\usepackage{dcolumn}% Align table columns on decimal point
\usepackage{bm}% bold math
\usepackage{float}
\usepackage{sidecap}
\usepackage{color}
\usepackage{epstopdf}
\graphicspath{{Pictures/}}

\begin{document}

\title{Energy exchanges between atoms with a quartz crystal $\mu$-balance}
  
\author{M. Guevara-Bertsch$^{1}$, A. Chavarr\'ia-Sibaja$^{2,3}$, A. God\'inez-Sand\'i$^{2,3}$, and O.A. Herrera-Sancho$^{2,3,4}$}

\address{$^1$ Institut f\"ur Quantenoptik und Quanteninformation,\"Osterreichische Akademie der Wissenschaften, Technikerstr. 21a, 6020 Innsbruck, Austria}
\address{$^2$ Escuela de F\'isica, Universidad de Costa Rica, 2060 San Pedro, San Jos\'e, Costa Rica}
\address{$^3$ Centro de Investigaci\'on en Ciencia e Ingenier\'ia de Materiales, Universidad de Costa Rica, 2060 San Pedro, San Jos\'e, Costa Rica}
\address{$^4$ Centro de Investigaci\'on en Ciencias At\'omicas Nucleares y Moleculares, Universidad de Costa Rica, San Jos\'e 2060, Costa Rica}

\date{\today}%
\vspace{10pt}

\begin{abstract}
 
We propose an experimental method to fully characterize the energy exchange of particles during the physical vapor deposition process of thin surface layers. Our approach is based on the careful observation of perturbations of the oscillation frequency of a Quartz Crystal $\mu$-balance induced by the particles interaction. With this technique it is possible to measure the momentum exchange of the atoms during the evaporation process and determine the ideal evaporation rate for an uniform energy distribution. We are able to follow the desorption dynamics of particles immediately after the first layers have been formed. These results are in close relation to the surface binding energy of the evaporated material, they offer a better control to obtain the desired properties of the thin surface layer. We applied our technique to investigate the physical vapor evaporation process for diverse elements, usually implemented in the development of film surface layers, such as Cu, W, Au, Gd and In, and confirm that our results are in agreement with measurements done previously with other techniques such as low temperature photo luminescence.

\end{abstract}

\maketitle
%%%%%%%%%%%%%%%%%%%%%%%%%%%%%%%%%%%%%%%%%%%%%%%%
\section{Introduction}
\label{introduction}

"Never measure anything but frequency!" was the motto of Arthur Schawlow~\cite{hansch2006einstein}, referring to the physical quantity that can be measured with the highest precision and accuracy. Through the measurement of frequency we can indirectly measure any other quantity and inherit, in the process, the same degree of exactitude. By solely measuring the perturbations of the frequency of a given oscillator, in a controlled environment, we can fully describe the electromagnetic field that surrounds him, the mass of the particles that hit him, and even  observe relativistic effects or detect gravitational waves~\cite{rosenband2007observation, nicholson2012comparison, itano2000external,guggemos2015sympathetic,kolkowitz2016gravitational}. The latter basically tells us basically that in order to understand and comprehend the laws of nature, we are only limited by how well we can measure frequency. As it is clearly stated by Arthur Schawlow, this fundamental principle has the advantage that it can be applied to measure any quantity and implemented to study any environment. Therefore we propose the application of this principle as a tool to fully characterize in situ and in vivo the physical vapor deposition process of thin surface layers.

The first layers of any material are the bridge of communication between the bulk and the environment that surrounds it. These layers have a key role in the comprehension and control of the properties of any material, see for example \cite{guevara2016detection}. Film surface layers have an incredible variety of roles: depending on their inner properties they can completely change the optical signature of a material and act either as an anti-reflective coating or, in contrast, as a perfect mirror~\cite{aspnes1982optical}; they are one of the key elements of biological cells in our body~\cite{kasemo2002biological}, regulating all fluid and gas exchanges; and finally they are now playing a crucial part in the development of magnetic storage materials~\cite{freund2004thin}. This is why the past decades have witnessed incredible technological development focused only in the properties and deposition techniques of thin surface layers~\cite{ pamplin2017molecular, moylan2003new,mattox2010handbook}. 

The central techniques in the synthesis of thin surface layers are based in physical vapor deposition (PVD). During PVD processes, a cloud of hot particles in a gaseous state is ejected into a designated substrate~\cite{reichelt1990preparation}. With the advent of nanotechnology, the control of the characteristics of thin surface layers has taken a more and more central part in the exploration of new physical effects, see for example Refs.~\cite{schmidt2001nanotechnology, beyertt2005optical, saas2006exciton}. Therefore the degree of control of all the particle dynamics during the evaporation processes determine the properties of the deposited surface layer~\cite{jasik2009influence}. Different techniques have been developed to observe the dynamics and interaction of the particles when they are in the gaseous state, such as for example time of flight experiments or Doppler shift methods~\cite{hagena1968time, phillips1998nobel, guggemos2015sympathetic}. They require, however, the implementation of expensive and complex equipment that is not always present in regular laboratories dedicated to surface science. Inspired by Arthur Schawlow's philosophy, we propose a new method to characterize the energy exchange of the particles during the evaporation process through the simple observation of the frequency of oscillation of a quartz crystal $\mu$-balance (QCM). 

The QCM is commonly used during surface film evaporation to measure the number of layers deposited. The frequency of oscillation of the QCM is constantly monitored during the evaporation: when the atoms are deposited on the surface, the frequency of oscillation changes consequently and allows us to easily determine the deposited mass. Our technique proposes to investigate the perturbations
of the frequency of oscillation of the QCM during the deposition process in even more detail. In this paper, we show how through the measurement of these perturbations we can determine the momentum exchange of the particles when they hit the surface, observe carefully the desorption of the atoms from the surface immediately after deposition and as a consequence determine the surface tension of the material. 

In the following sections we are going to present first, in section~\ref{section:exp}, our experimental set up. In section~\ref{section:QCM}, we will present a brief introduction of the principle of operation of the QCM and how we implemented the oscillator to detect environmental perturbations. Following this, in section~\ref{section:momentum}, we describe  an experimental method to investigate the momentum exchange of the atoms during the deposition. In section~\ref{desorption} we apply this technique to observe the energy dynamics that occur due to the desorption of the atoms from the thin surface layer. Finally, in section~\ref{conclusion}, we summarize and explain how the techniques described can be implemented to fully characterize the properties of a deposited surface during the evaporation process.

\section{Experimental setup} \label{section:exp}
%%%%%%%%%%%%%%%%%%%%%%%%%%%

\begin{figure}[h]
\begin{center}
\includegraphics[height=6cm]{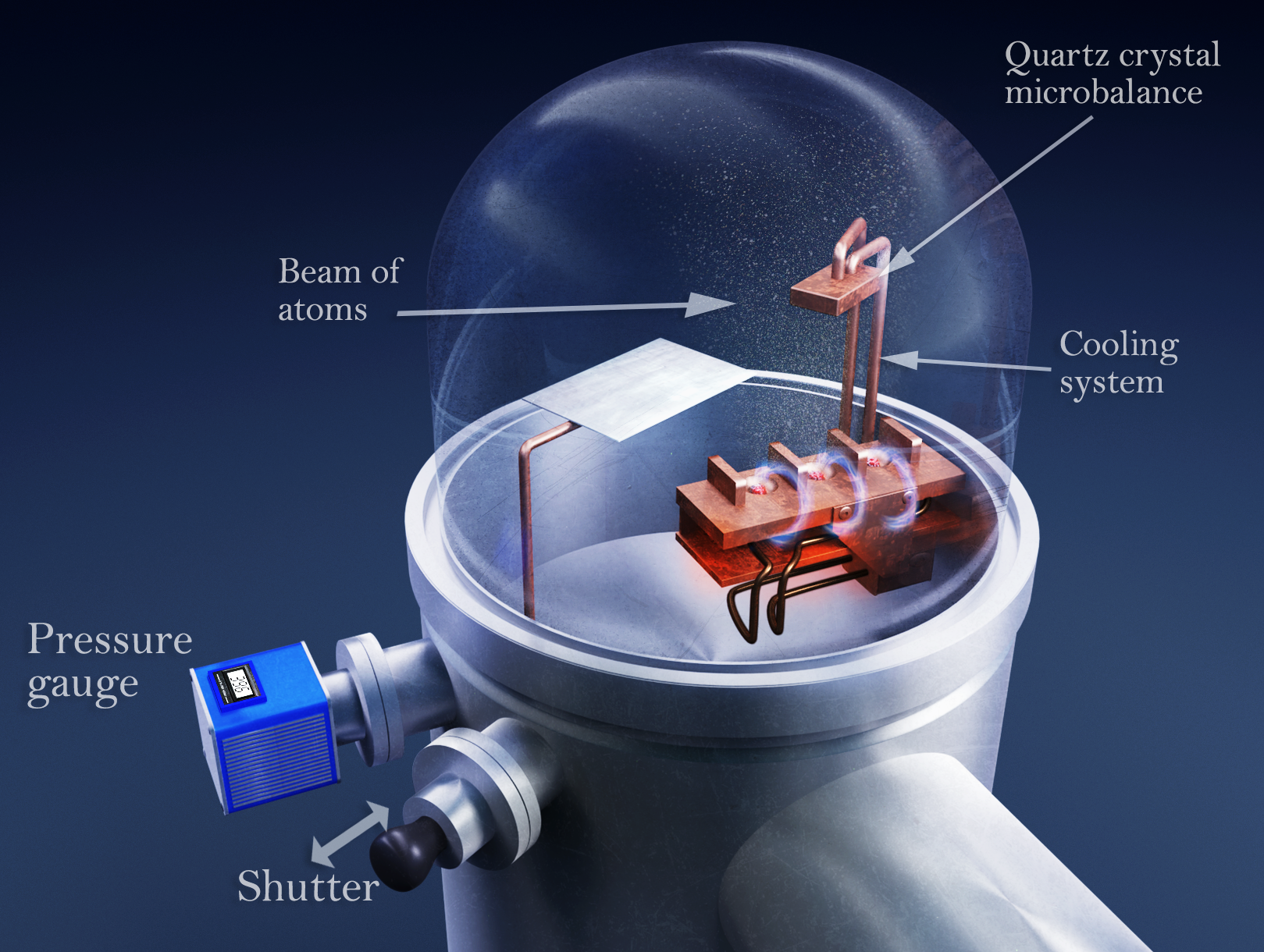}
\caption{(Color online )Artistic visualization of our experimental setup. A quartz crystal $\mu$-balance (QCM) 
is situated at about 28\,cm directly above the position of the target material (e.g. Cu, W, Au, Gd and In). A water cooling system allows us to stabilize the temperature of the QCM during the evaporation. The shutter (see lower part of the apparatus) is manually controlled from the outside in order to block at will the beam of atoms which interact with the surface of the QCM. During all experiments carried out in this investigation the system is maintained at high vacuum pressure of around 10$^{-7}$\,Torr ($10^{-5}$Pa).}
\label{Fig: Montaje}
\end{center}
\end{figure}
%%%%%%%%%%%%%%%%%%%%%%%%%%%
The experimental setup is illustrated in Fig~\ref{Fig: Montaje}. In order to characterize and monitor the energy transfer of the atoms during the evaporation and deposition process we use a quartz crystal $\mu$-balance (QCM) coupled to a thickness monitor model Maxtek TM-400 and a frequency meter model Keysight 54 220A. This system allows us to observe any frequency fluctuations with a resolution of approximately $10$\,mHz. The temperature of the quartz crystal is monitored with a J type thermocouple located in the housing of the QCM, as close as possible of the crystal surface, and is kept constant by means of a water cooling system ($\Delta$T$\approx 1$\,K).A small amount of the solid element (e.g. Cu, W, Au, Gd and In) is placed in a crucible at a distance of about 28\,cm below the QCM. In favor of changing the phase of the material from a condensed phase to a vapor phase, i.e. evaporation, we use an electron cannon model Thermionic 150-0040. The latter allows us to generate an electron flow with variable emission current to reach the evaporating temperature of the desired material as well as a tunable evaporation rate. A shutter near the QCM surface (see Fig.\ref{Fig: Montaje}) allows us to immediately stop the interaction between the beam of atoms and the QCM. In order to avoid any environmental  contamination, the system is mounted inside a high vacuum apparatus modified from that of Ref.~\cite{clark2014polycrystalline} with the capacity to maintain pressures in the range of roughly 10$^{-7}$\,Torr ($10^{-5}$\,Pa). The pressure is constantly monitored by a pressure gauge model 354 MICRO-ION and maintained stable in the order of $~10^{-7}$\,Torr ($10^{-5}$\,Pa) during all measurements reported below. 

%%%%%%%%%%%%%%%%%%%%%%%%%%%
\begin{figure}[h]
\begin{center}
\includegraphics[height=8cm]{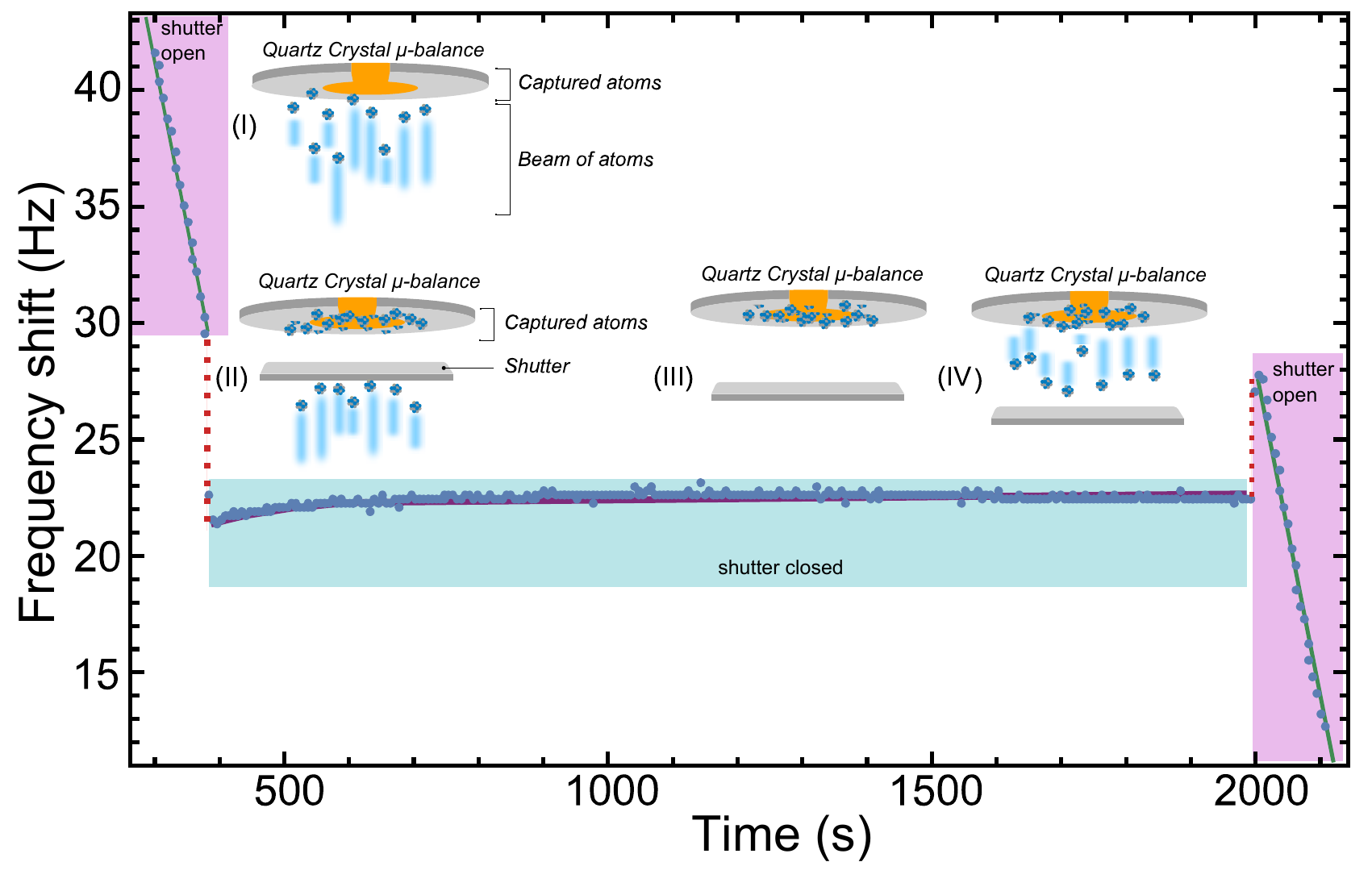}
\caption{(Color online) Experimental data showing the time evolution of the energy transfer during the evaporation process, with two events of opening and closing the shutter. In section~(I): a constant beam of atoms is interacting with the quartz crystal reducing the frequency of oscillation; in section~(II) the beam of atoms is abruptly stopped with the shutter. The observed change in frequency $\Delta f$ (see Eq.~\ref{eq:freqmass}), of about 9 Hz, corresponds to the momentum transfer of the atoms hitting the quartz crystal; in section~(III) the shutter is kept closed for a time t$\approx1600$\,s. In section~(IV) we observe an increase of the frequency due to desorption of atoms from the surface. Finally, the shutter is reopened, a jump in the frequency is again observed due to the momentum transfer of the atoms hitting the quartz crystal and the QCM continues decreasing as described by Eq.\ref{eq:Sauerbrey} from $t\geq 2000$\,s.}
\label{fig:energyfreq}
\end{center}
\end{figure}
%%%%%%%%%%%%%%%%%%%%%%%%%%%

\section{Quartz Crystal $\mu$-balance}
\label{section:QCM}

The operating principle of the QCM is the following: a crystal oscillator is used to detect any change of frequency of a piezoelectric quartz crystal due to the perturbations caused by the deposition of small amount of mass with high precision~\cite{Lu-2012}. Typically, QCM systems, as indicated by their name, are used in high precision and low order of magnitude mass measurements, such as the growing of thin films where it is necessary to determine the mass up to levels of $10^{-12}$\,kg~\cite{Pettersson-1919, lu2012applications}. From the physical point of view, by means of the Sauerbrey Eq.~\cite{Sauerbrey-1959}, the variation of the deposited mass is related to a decrease in the resonant oscillation frequency of the crystal, $f_{0}$ (usually between 6-9 MHz), as follows:

\begin{equation}
\Delta f_{ev}=-\frac{2f_o^{2}\Delta m}{A\sqrt{\rho\mu}},
\label{eq:Sauerbrey}
\end{equation}

where $\Delta f_{ev}$ is the variation of the oscillation frequency (units in Hz) of the crystal during the deposition, $\Delta m$ the total mass (units in g) detected by the balance (explained in more detail in section~\ref{section:momentum}), $A$ the surface (units in cm$^{2}$) of the deposited film, $\rho$ the quartz density (with a constant value of $2.648$ g$\cdot$cm$^{-3}$) and $\mu$ the shear modulus of the crystal (with a constant value of $2.947\cdot 10 ^{11}$g$\cdot$cm$^{-1}\cdot$s$^{-2}$); the reported values for $\rho$ and $\mu$ correspond to our experimental QCM described by the manufacturer. Equation~\ref{eq:Sauerbrey} is commonly used in surface science to determine the thickness rate during evaporation of a material during the growth of thin films, see for example Refs.~\cite{kurosawa2000detection,leskela2002atomic,yousfi2001atomic}. Usually, during PVD experiments, such as molecular beam epitaxy (MBE), the QCM is placed as close as possible to the substrate sample in order to mimic, in an appropriate way, the "real" thickness on the surface of the substrate. The thickness of the film in the substrate is correlated to the change in frequency of the QCM, allowing us to control the deposition rate. As the atoms are deposited on the surface of the crystal the oscillation frequency,$f_{ev}$ decreases consequently. As explained in the seminal papers that developed and studied the QCM applications~\cite{kasemo1978quartz,kasemo1979kinetics}, through the careful observation and measurement of the perturbations of the QCM oscillation frequency during the evaporation process, it is possible not only to extract the deposition rate but, even more interesting from the dynamics point of view, to fully characterize the energy exchanges between atoms. In the following sections we are going to present how to implement the QCM to measure the momentum exchange of the atoms as well as the desorption mechanism during the deposition process to characterize and control the properties of the formed surface layer.

\section{Characterization of the momentum exchange}\label{section:momentum}

	We start by focusing our attention at the instant when the atoms hit the surface of the crystal. If we assume that only the atoms that are ejected from the crucible interact with the QCM, depending on the electric current applied to the electron cannon a number $N$ of atoms per second hit the crystal surface with a mean speed $v$. The cumulative transfer of momenta  generated by the collision of these atoms with the crystal surface generates a constant negative perturbation of the oscillation frequency of the QCM. As a consequence, the oscillation frequency of the crystal $f_{ev}$ is equal to: $f_ {ev}= f_{a} + \Delta f$, see Fig.~\ref{fig:energyfreq}, section II, where $f_{a}$ corresponds to the frequency variation due to the deposited mass and $\Delta f$ to the frequency offset generated by the momentum exchange. From the point of view of the QCM this constant perturbation is perceived as a "virtual negative mass". Therefore, the amplitude of this perturbation is equal to the total energy exchange of the "hot" beam of atoms that arrive at the surface of the crystal with the "cold" atoms that are already deposited in the surface. As described in the references~\cite{lu2012applications,kasemo1978quartz}, through the measurement of this "virtual mass" we can extract not only the mean speed $v$ at which the beam of atoms is hitting the surface of the QCM but also the atoms momentum exchange. To obtain a direct relation between the atom\'s momentum exchange and the frequency offset, we start by the defining the total mass perceived by the QCM, $\Delta m$, as:

\begin{equation}
\Delta m = m_p+ Nm_a-m_v,
\label{eq:masa}
\end{equation}

where $m_{p}$ is the mass of the quartz crystal, $m_{a}$ the atomic mass of the evaporated element and $m_{v}$ the virtual mass. The offset frequency $\Delta f$ corresponds to the difference of frequency before and after the shutter is closed as illustrated in Fig.~\ref{fig:energyfreq}, section II. As we measure $\Delta f$ in our apparatus, it is convenient to rearrange Eq.~\ref{eq:Sauerbrey} and Eq.~\ref{eq:masa} in order to obtain $\Delta f$ as a function of the virtual mass $m_{v}$:

\begin{equation}
\Delta f= \frac{2f_{o}^{2} m_{v}}{A\sqrt{\rho\mu}}.
\label{eq:freqmass}
\end{equation}

 The variation of the total momentum exchange $q$ of the atoms when they hit the crystal surface is equal to the exerted force of the atoms on the surface given by:

\begin{equation}
m_v g= \frac{dp}{dt}=q,
\label{eq:10}
\end{equation}

where $g$ is the gravitational acceleration constant and $p$ the particle momentum. As a consequence, by inserting $m_v$ from Eq.~\ref{eq:freqmass}, we can determinate $q$:

\begin{equation}
q=\frac{Ag\sqrt{\rho\mu} \Delta f}{2f_o^{2}}.
\label{freqvsrapidez}
\end{equation}

	Therefore by means of measuring $\Delta f$ for different scenarios (for example for different evaporation rates, or different species of atoms) we can easily characterize $q$. Experimentally this offset frequency $\Delta f$ corresponds to the difference between the frequency before and after the beam of atoms is blocked by a shutter, as presented in the section~II of Fig.~\ref{fig:energyfreq}. 
In order experimentally implement the scheme described above, we applied the following cycle:

\begin{enumerate}
	\item evaporation of the element with a fixed rate during $60$\,s,
	\item closing the shutter during $1600$\,s,
	\item opening the shutter,
	\item increase the evaporation rate, and finally 
	\item repeat the cycle.
\end{enumerate}

	For each of these cycles the frequency shift $\Delta f$ is measured, as it is shown in section II of Fig.~\ref{fig:energyfreq}, and the momentum exchange $q$ extracted by using Eq.~\ref{freqvsrapidez}. 
	
\begin{figure}[H]
\begin{center}
\includegraphics[height=6cm]{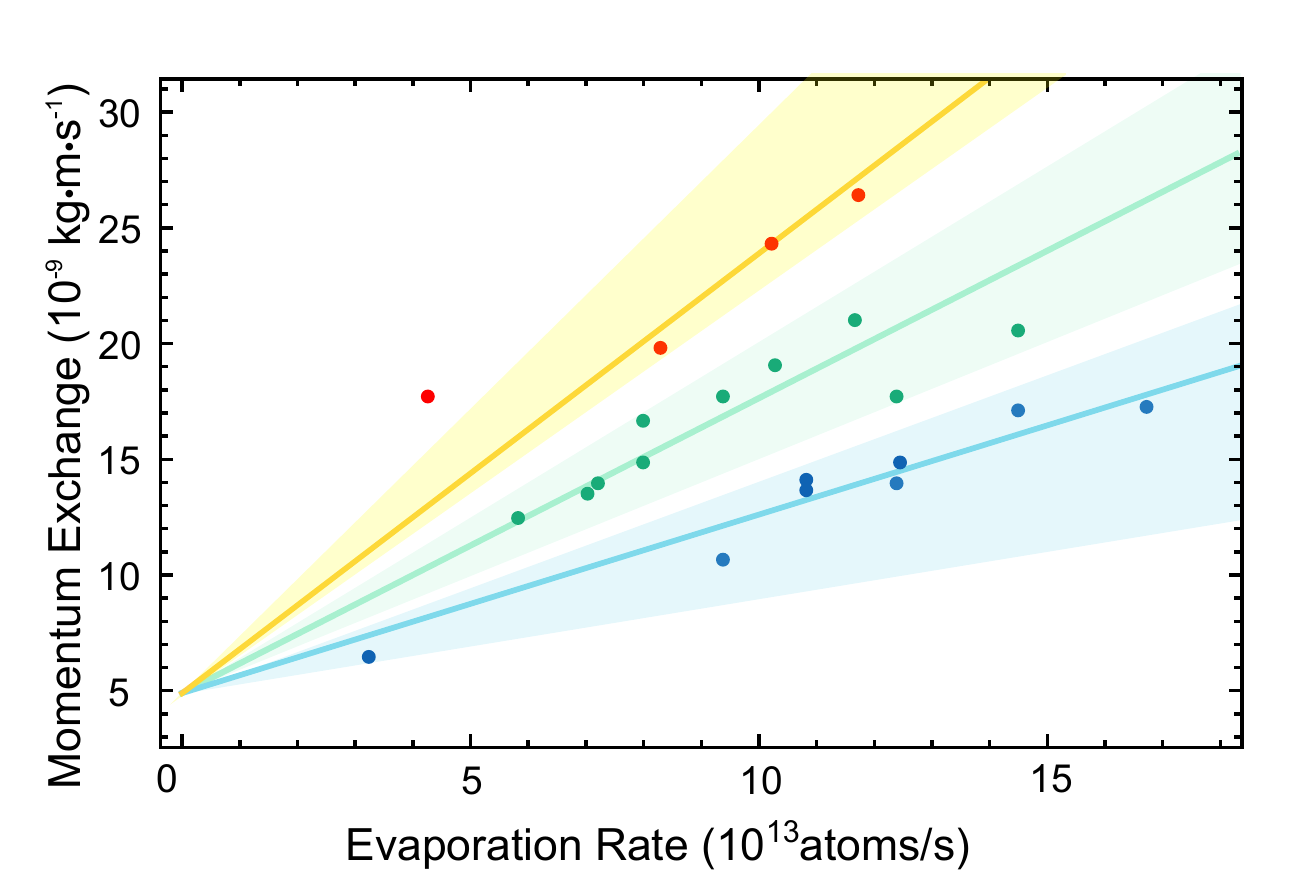}
\caption{(Color online) Variation of the momentum exchange due to the interaction of the evaporated atoms with the ones present in the surface layer. The measured data was separated in three classes (shaded colors), according to the average speed of the atoms, extracted through the application of a linear fitting of the data points: yellow ($625\pm65$\,$m/s$)m/s; green ($405\pm78$\,$m/s$) m/s and blue ($248\pm46$\,$m/s$)}.
\label{fig:velocityvstime}
\end{center}
\end{figure}

	As an experimental demonstration we applied the technique described above to characterize the momentum exchange of a beam of W (Tungsten) atoms, during the evaporation of a thin surface layer. For a series of different $N$ we measured $q$. Figure~\ref{fig:velocityvstime} shows the variations of the momentum $q$ as a function of the evaporation rate. The momentum exchange increases, as expected, proportionally to the evaporation rate $N$. As it is well known, the atoms that are randomly ejected from the crucible do not all have the same mean speed. As described for similar measurements in Ref.~\cite{meyer1981thermalization}, once the temperature of evaporation is reached the evaporated atoms acquire different speeds, following approximately the Maxwell distribution. This speed distribution is, as expected, reflected in our measurements of $q$: Fig.~\ref{fig:velocityvstime} shows a spread distribution of the momentum exchange as a function of the evaporation rate. We observe that we can group this "spreading" in three main sections, distinguished by the different colored shadings in Fig.~\ref{fig:velocityvstime}. The main velocity of each section is extracted from the slope of a linear fitting applied to the data of each section. According to Fig.~\ref{fig:velocityvstime}, the velocity distribution of the hot W atoms can be separated in three main different velocity classes: $625\pm65\,m/s$ (yellow); $405\pm78\,m/s$ (green) and $248\pm46\,m/s$ (blue). These values are in agreement with the theoretical velocity distribution obtained from the vapor pressure measured during the experiment of about $10^{-7}$\,Torr which corresponds to a temperature of approximately $2300~^\circ$C, as described in Ref.~\cite{Roth-2012}.

%%%%%%%%%%%%%%%%%%%%%%%%%%%

%%%%%%%%%%%%%%%%%%%%%%%%%%%
\textbf{}

	During the deposition process, the energy exchange between the atoms at the moment they are in contact with the surface affects the properties of the surface layer. A. Jasik and his group explain how the growth conditions affect directly both the interface smoothness and the overall layer quality~\cite{jasik2009influence}. The same investigation also points out how the main difficulty comes from the necessity of controlling all variables involved during the evaporation process, since it depends on a particular apparatus and experimental conditions. Through the measurement and characterization of the distribution energy exchange of the atoms during the deposition process, we can define the ideal conditions of the evaporation process. We apply our measurement technique to develop an experimental method in situ to determine the optimum deposition rate for a given material. This value corresponds to the point where the energy exchange between the atoms reaches maximum efficiency, meaning a point where most of the particles arrive at the surface with the same energy. Therefore all the velocity classes contribute coherently to reach the maximum point. As mentioned before, accordingly to the current applied to the electron gun a hot cloud of $N$ atoms per second is emitted from the crucible. Before they reach the target, the highly energetic atoms in the cloud are going to interact and exchange energy with each other in a random manner, as explained in Ref.~\cite{huang1987statistical}. Following a classical approximation, the number of collisions between atoms depends on the mean free path they can travel in the cloud. The latter means that the mean free path depends, among many other variables, on the mass and the number of atoms $N$ in the cloud. For a fixed $N$ with higher masses we expect a higher number of collisions for the same volume, resulting in more abrupt energy exchanges. On the other hand, for a fixed mass, for low values of $N$ we expect small variations of the momentum exchange, since the amount of collisions is relatively small in comparison with the previous case. As the number of atoms increases, the number of collisions increases accordingly until it reaches a value where the energy exchange between all particles per second is such that it accomplishes a resonance point (maximum energy interchange or maximum efficiency) of the whole cloud of atoms. The optimum value for the evaporation rate corresponds to the point at which the  energy variations are minimized and therefore the cumulative momentum exchange of the system is maximized. 
	
\begin{figure}[H]
\begin{center}
\includegraphics[height=6cm]{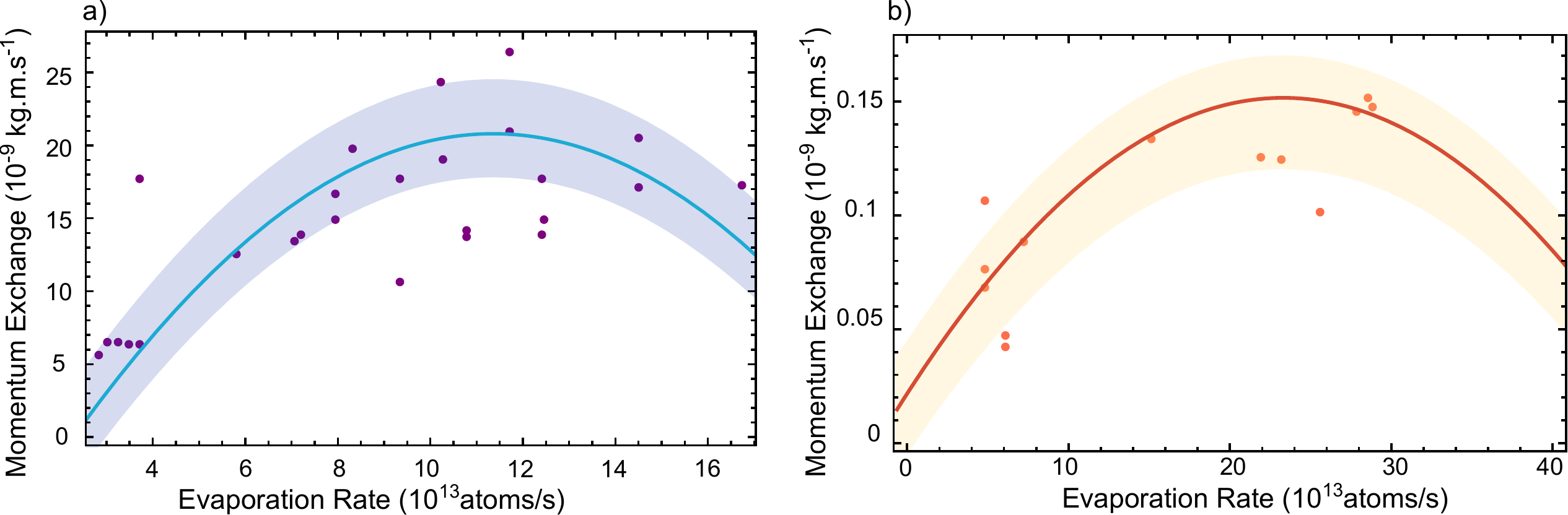}
\caption{(Color online)Distribution of the energy exchange of the atoms  as a function of deposition rate during the evaporation. Figure~a) corresponds to the data taken for W atoms and figure~b) for Cu atoms. In both cases the solid line is a Gaussian fit of the total ensemble and the shading illustrates the maximum dispersion of the data from the Gaussian fit.}
\label{eficiencia}
\end{center}
\end{figure}

		In order to characterize our experimental method we compare the energy distribution of two elements with different atomic masses: Cu ($63.5$\,u) and W ($183.4$\,u). As expected, the maximum value of the distribution of energy is one order of magnitude higher in the case of W than for the case of Cu atoms, as it can be see in Fig.~\ref{eficiencia}. From these measurements we can also determine that the optimum rate to obtain an uniform energy distribution for Cu is either at very low rates of approximately $3\times 10^{13}$ atoms$/$s or at high values of approximately $16\times 10^{13}$~atoms$/$s and for the W is also for low values of approximately $5\times 10^{13}$~atoms$/$s or for high values of $50\times 10^{13}$~atoms$/$s. These results are in accordance with a similar observations, reported by A. Jasik in Ref.~\cite{jasik2009influence}, where the relation between the growth rate and the quality of structures by MBE is studied. In this investigation the quality of the newly deposited structure is studied by means of low temperature photo-luminescence (PL). The data of the variations of the PL as a function of the deposition rate presented in Fig.~1b of Ref.~\cite{jasik2009influence} can be fitted to a Gaussian fit in a similar matter as our experimental observation (see Fig.~\ref{eficiencia}). To corroborate this relation the residuals from a gaussian fit for both data sets where calculated. In both cases the residuals tend to zero which confirms the proposed gaussian model, for example, in the data set obtained with the PL experiment the highest residual is on the order of $10^{-2}$ and for the data set obtained with the QCM is on the order of $10^{-1}$. The measurements with the QCM are performed in a high temperature conditions. The difference of one order of magnitude between the residuals is explained by the measurement difficulties that arise in such a noisy environment. Nonetheless through the simple observation of the variations of the oscillation frequency of the QCM we are able to characterize the deposition process and obtain similar results than with more sophisticated systems as low temperature PL experiments.

%%%%%%%%%%%%%%%%%%%%%%%%%%%

\section{Characterization of the desorption mechanism}\label{desorption}

\begin{figure}[H]
\begin{center}
\includegraphics[height=12cm]{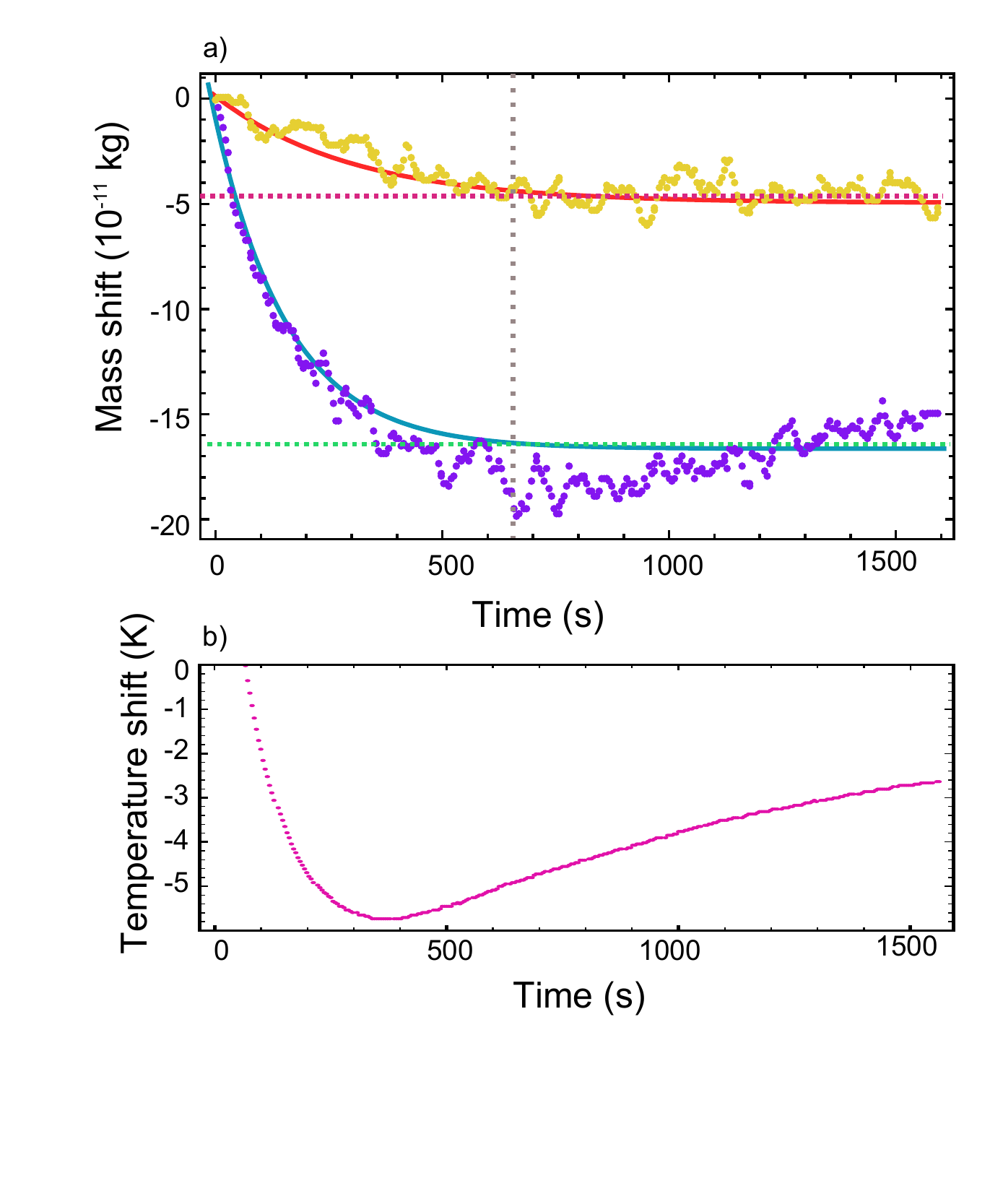}
\caption{(Color online) Characterization of the desorption process. a) Variation of the atoms present in the QCM as a function of time once the shutter is closed. The yellow represents Au atoms and the purple represents W atoms. The solid line, red for Au atoms and green for W atoms, corresponds to the respective exponential fit of the mass decay. b) Is the variation of temperature exactly at the same time that the data in a) was taken.}
\label{desprendimiento}
\end{center}
\end{figure}

	In this section we consider closely the energy exchange that occurs between atoms present in the surface layer once they have been deposited. At this point, before the thermalization is reached, the freshly deposited surface atoms are still exchanging energy between themselves and the environment. With the increasing interest in electronic cooling and heat transfer in nanotechnology, the understanding of  energy transfer mechanisms during the thermalization of the surface can be of crucial importance \cite{kim2007spray}. During this step, the final properties of the surface layer are going to be established which makes it a perfect time scale to analyze and characterize the deposition process. In order to carry out this investigation we focus our attention in the variations of the frequency of the QCM in the following 1500\,s after the shutter has been closed, as it is illustrated in Section III, Fig.~\ref{fig:energyfreq}. We observe an increase of the frequency of oscillation which corresponds to a decrease of the number of atoms previously deposited at the surface. In order to quantify this effect, we can estimate the mass variation from the frequency of the QCM by using Eq.~\ref{eq:Sauerbrey} as: $\Delta m=-\frac{2f_o^{2}}{A\sqrt{\rho\mu} \Delta f_{ev}}$. Figure~\ref{desprendimiento} a) presents the mass variations measured during this time scale for the case of Au (orange data) and W(blue data) atoms. The figure also clearly shows an exponential decay of the mass present in the surface, that reaches, in both cases, a threshold after a time of approximately 700\,s. This tendency is in agreement with the characterization of the desorption process with a QCM presented in the seminal work by Kasemo~\cite{kasemo1979kinetics}. Therefore, our measurements imply the presence of a desorption mechanism triggered once the shutter is closed. 
	%%%%%%%%%%%%%%%%%%%%%%%%%%%

%%%%%%%%%%%%%%%%%%%%%%%%%%%

	In order to explain the detachment of the newly deposited atoms from the surface, we considered the surface binding energy of both deposited materials. A transfer of energy sufficiently large to the deposited atoms could cause a break of the surface binding energy and explain the desorption mechanism. The surface binding energy is the first link between the oncoming energetic atoms and the ones present in the surface of the crystal. As described by Rendulic and his group,~\cite{rendulic1994adsorption}, this bond starts to be formed by the atoms at the instant they are in contact with the surface to minimize their energy. Desorption dynamics are usually related to temperature effects. As it is naively expected, a rise of the temperature leads to an increase of the probability that the atoms return to the gas phase~\cite{redhead1962thermal, king1975thermal}. In their research, Rendulic et al ~\cite{rendulic1994adsorption} present the development of several procedures to characterize the adsorption and desorption dynamics as a function of external temperature variations. Nonetheless, our measurements of the variation of temperature in the proximity of the QCM, presented in Figure~\ref{desprendimiento} b) show, during the same time, a decrease of approximately 5~K at the instant the shutter is closed. From these measurements we can exclude any external temperature variation to be at the origin of the desorption of the atoms from the surface. Any chemical reaction can also be excluded, since the chamber is maintained at high vacuum environment. That leave us with the inner temperature of the hot atoms that just arrived to the surface. From the measured vapor pressure the temperatures of the W and Au atoms at the moment they reach the surface are $T_{W}=2300\,^{\circ}$C and $T_{Au}=987\,^{\circ}$C, respectively~\cite{Roth-2012}. The stored energy of the hot atoms is going to be released simultaneously to the environment and to the atoms present in the surface until a thermal equilibrium is finally reached. Before thermalization is reached, an equilibrium between the solid and gaseous phases is established. Under these conditions the rate of vaporization is equal to the rate of condensation according to Ref.~\cite{Roth-2012}. At the instant the shutter is closed, until the termalization is finally reached, part of the atoms present in the surface return to the gaseous phase. 
	
\begin{figure}[H]
\begin{center}e
\includegraphics[height=6cm]{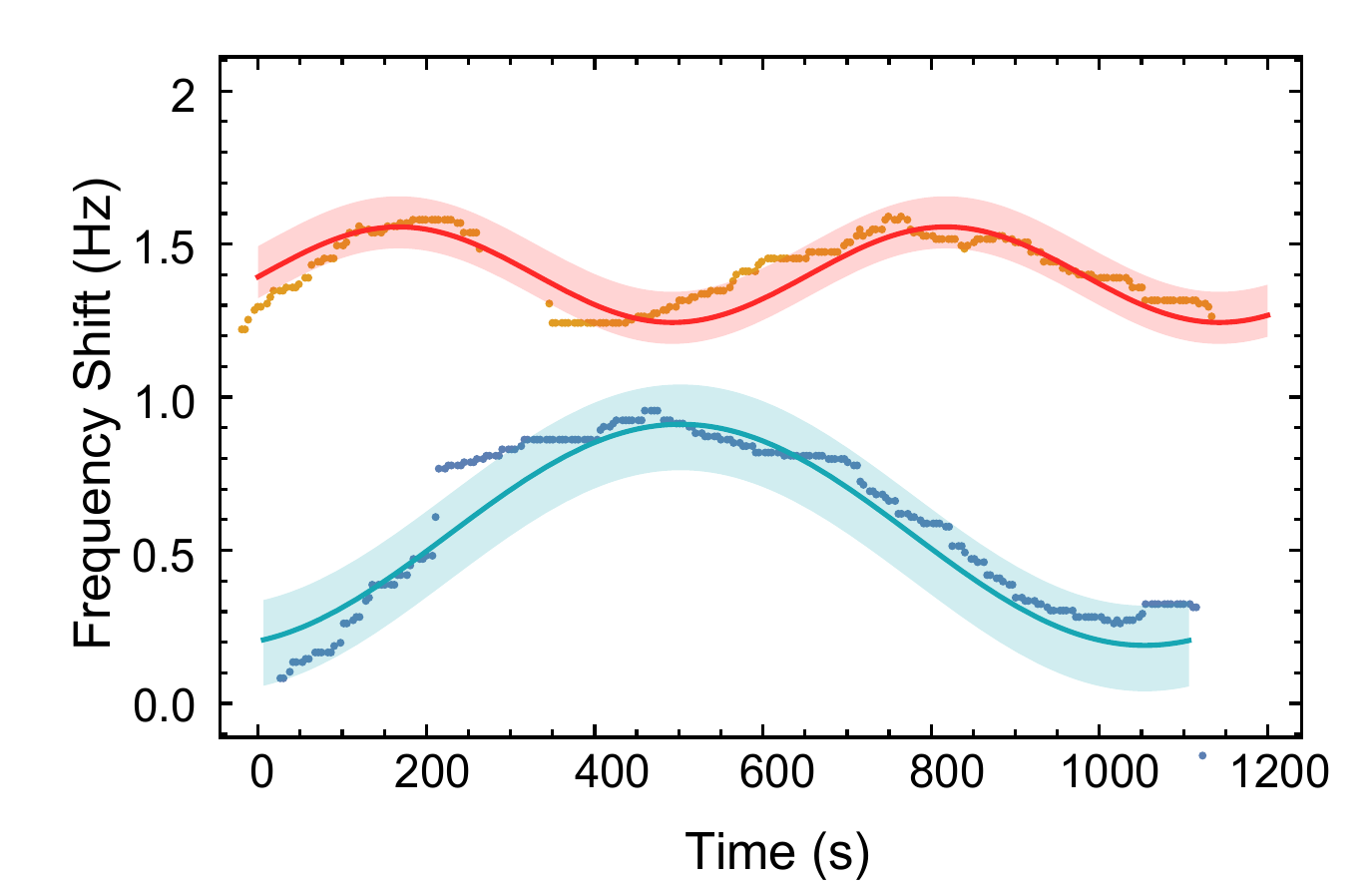}
\caption{(Color online) Frequency modulation of the QCM. The orange points correspond to the data taken for Au atoms and in purple for W atoms. The solid line in both cases, red for Au and blue for W atoms, corresponds to a sinusoidal fit applied to the respective set of data. The shading illustrates the dispersion of data from the sinusoidal fit.}
\label{oscilaciones}
\end{center}
\end{figure}

	As shown in Fig.~\ref{desprendimiento}, during the time $t_{sat}$, a total mass of W: $m_{W} = 16\cdot 10^{-11}$~kg and a total mass of Au: $m_{Au} = 5\cdot 10^{-11}$~kg is detached from the surface. The time at which the threshold is reached corresponds to the thermalization of the surface thin layer. The energy absorbed from the hot atoms during deposition is mainly absorbed by the QCM housing and transferred to the cooling system during the time $t_{sat}$. During this time part of this energy is also absorbed by the newly deposited atoms which are at the origin of the desorption process. The mass of atoms detached, in the same period of time, for each material is dependent on how strongly they are attached to the other atoms, meaning on how big their surface binding energy is. The surface binding energy of W and Au is $E_{bindingW}=11.75$~eV and $E_{bindingAu}=4.13$~eV, respectively~\cite{kudriavtsev2005calculation,yang2014atomic}, as we would like to compare in a dimensionless way, the behavior of the physical quantities involved in our system, we notice that the ratio of this two energies ($\frac{E_{bindingW}}{E_{bindingAu}}$) is approximately equal to $2.8$ which corresponds closely to the ratio of the detached masses ($\frac{m_{W}}{m_{Au}}$). Fom these measurements we have a clear relation between the surface binding energy and the number of atoms that are detached from the surface of the thin layer. We looked even closer at the frequency variations of the QCM after closing the shutter and observed a modulation of the oscillation frequency. As it is presented in Fig.~\ref{oscilaciones} once the atoms are detached, after we subtracted the exponential fit shown in Fig.~\ref{desprendimiento} a), we observe a frequency modulation of the QCM. The period of oscillation is for Au approximately $\tau_{Au}=800~s$ and for W about $\tau_{W}=2200~s$. Bearing in mind that what interests us is the comparison of dimensionless quantities, we simply calculated the ratios for these periods and surprisingly enough we obtained a value that is roughly equal to the ratio of surface binding energies. The observed modulation is originated by the detachment of the atoms from the surface. The strength of the binding energies determines the frequency of the modulation in the same way that the Hooke constant affects the frequency of oscillation of a damped spring. These measurements open the door to a new way to characterize in situ the properties of newly deposited thin surfaces. By means of only a QCM we could acquire a full knowledge and control of the fundamental properties of the deposited surface.

%%%%%%%%%%%%%%%%%%%%%%%%%%%

\section{Conclusion}\label{conclusion}

In conclusion, we have demonstrated how the QCM can be implemented as an energy transfer sensor to detect the dynamics between particles during the physical vapor deposition of thin surface layers. Confirming the motto of Arthur Schawlow, our measurement technique is based in the careful observation of the perturbations of the oscillation frequency of the QCM. We are able to sense the particle interaction in the vapor cloud before, during and after they have been deposited at the substrate. We investigated the momentum exchange of the particles with the QCM and extracted the average speed distribution of the particle cloud. From these measurements we determined the optimum deposition rate that allows the achievement of a uniform energy distribution during the physical vapor deposition process. Our results support the theoretical predictions as well as similar observations reported by A. Jasik using other techniques in rather complicated measurement systems such as low temperature photo luminescence in rather complicated measurements~\cite{jasik2009influence}. We also studied the bounding formation of the particles in a freshly formed layer. We noticed, during the following seconds of the deposition, a detachment of particles. From the careful analysis of the variations of the frequency, we detected the dynamics of the desorption mechanism and noted a clear dependence with the surface binding energy. 

With this set of sensing techniques, we propose a simple and easy to implement method to monitor the complex energy interactions during the physical vapor deposition of thin surface layers. The main advantage of our measurement approach is that it only requires the implementation of a QCM to the experimental setup. Theoretical studies suggest that in order to mediate the atomic assemble to create desired atomic structure it is necessary to reach a better understanding of the complexity of the vapor deposition. Diverse models have been suggested where it is necessary to compute the velocity and position of every atom in the system so that they can be compared to the highest lattice vibration and used to predict the molecular dynamics~\cite{wadley2001mechanisms}. Our method offers an accessible and simple solution to have a direct overview of the interplay of these complex variables and study with a deeper understanding any physical vapor deposition system such as thermal evaporation, sputtering, pulsed laser deposition techniques and molecular beam epitaxy among others~\cite{mahan2000physical}.

\section{Acknowledgments}
We would like to thank V\'ictor Rodr\'iguez Araya for his expert technical support in the construction of the experiment. We also want to acknowledge the helpful assistance of the graphic designers Felipe Molina Guti\'errez and Allan Fonseca for their effort in creating graphic art from our experimental apparatus. The authors are grateful for the support given by the Vicerrector\'ia de Investigaci\'on at the Universidad de Costa Rica to carry out this research work.

\section{References}

\bibliography{refs} 
\bibliographystyle{iopart-num}

\end{document}